\documentclass[runningheads]{llncs}
\usepackage[T1]{fontenc}
\usepackage{graphicx}
\usepackage{amsmath}
\usepackage{amssymb}
\usepackage{xcolor}
\usepackage[sectionbib,numbers]{natbib}
\usepackage{multirow}  
\usepackage{tabularx}
\begin{document}

\title{Enhancing Framingham Cardiovascular Risk Score Transparency through Logic-Based XAI}

\titlerunning{Enhancing FRS with Logic-Based XAI}
%
\author{Emannuel L. de A. Bezerra\inst{1}\orcidID{0009-0004-0195-9683}\and
Luiz H. T. Viana\inst{1}\orcidID{0009-0005-1509-6038} \and
Vinícius P. Chagas\inst{1}\orcidID{0009-0003-3871-640X}\and
Diogo E. Rolim \and
Thiago Alves Rocha\inst{1}\orcidID{0000-0001-7037-9683} \and
Carlos H. L. Cavalcante\inst{1}\orcidID{0000-0001-9395-8338}
}

\authorrunning{Bezerra et al.}
%
\institute{Instituto Federal do Ceará (IFCE), Maracanaú, Ceará Brazil\\
\email{\{emannuel.bezerra07, henrique.viana06, vinicius.peixoto.chagas61\}@aluno.ifce.edu.br}\\ 
\email{\{thiago.alves, henriqueleitao\}@ifce.edu.br}}
\maketitle              
\begin{abstract}
Cardiovascular disease (CVD) remains one of the leading global health challenges, accounting for more than 19 million deaths worldwide. To address this, several tools that aim to predict CVD risk and support clinical decision making have been developed. In particular, the Framingham Risk Score (FRS) is one of the most widely used and recommended worldwide. However, it does not explain why a patient was assigned to a particular risk category nor how it can be reduced. Due to this lack of transparency, our approach introduces a logical explainer for the FRS. Based on first-order logic and explainable artificial intelligence (XAI) fundaments, our method is capable of identifying the minimal set of patient attributes that are sufficient to explain a given risk classification (abduction). The explainer produces actionable scenarios that illustrate which modifiable variables would reduce a patient's risk category (counterfactual). 
We evaluated all possible input combinations of the FRS (over 22,000 samples) and tested them with our solution, successfully identifying important risk factors and suggesting focused interventions for each case. The results improves clinician trust and facilitates the wider implementation of CVD risk assessment by converting opaque scores into transparent, prescriptive insights, particularly in areas with restricted access to specialists.

\keywords{Framingham Risk Score \and Explainable Artificial \and Logic-Based \and Explanation
Intelligence \and Cardiovascular Disease.}
\end{abstract}
\section{Introduction}

Cardiovascular disease (CVD) is a critical public health issue, impacting millions of people worldwide and imposing a significant economic strain on healthcare systems \cite{kazi2024forecasting}. In 2017, the global mortality rate was approximately 55 million deaths, with cardiovascular diseases (CVD) accounting for 32\% of the total \cite{kazi2024forecasting}. Predicting the 
risk of developing CVD for individuals plays a crucial role in prevention, allowing for the early identification of those who may need more focused interventions \cite{arnett20192019}.

Many risk prediction models have been presented in the last few decades \cite{damen2016prediction}.  The Framingham Risk Score (FRS) is still one of the most widely validated instruments among them \cite{kasim2023validation}; the Brazilian Society of Cardiology's (SBC) Department of Atherosclerosis uses it as well \cite{precoma2019updated}.  Originally designed to predict the 10-year risk of coronary heart disease (CHD), the Framingham Heart Study (FRS) is based on a prospective study that was started in 1948 and has followed thousands of individuals \cite{wilson1998prediction}.  Later on, it was expanded to forecast a wider range of cardiovascular conditions \cite{d2008general}.

FRS employs sex-specific equations that integrate key risk factors: age, total cholesterol (TC), high-density lipoprotein cholesterol (HDL-C), systolic blood pressure (SBP), smoking status, and diabetes. Based on an individual’s score, FRS assigns them to low, moderate, or high risk categories to develop CVD over the next decade \cite{d2008general}.

Evidence from multiple cohorts demonstrates that FRS reliably identifies those at elevated risk in various populations - including samples from the general community \cite{d2008general}, multiethnic groups \cite{chia2015validation}, patients with diabetes or metabolic syndrome \cite{wilson2005metabolic}, and women with polycystic ovary syndrome \cite{amiri2025cardiovascular}.

Fortunately, most of the risk factors are modifiable, and timely intervention can reduce adverse outcomes and deaths, making the CVD risk prediction of individuals critical in the precise and efficient reduction of deaths.

Although FRS effectively supports clinical decision-making, it only reports a numerical risk score and category, leaving unclear which factors drove the classification and what specific changes a patient should make to lower their risk. This opacity can impact the confidence in the model's result and limits actionable guidance. 

Therefore, it is essential to highlight the individual risk factors that influenced the classification and to indicate which modifiable characteristics the patient should address to reduce their future risk. We address this gap by using first-order logic to build an abductive and counterfactual explainer for the Framingham Risk Score, thereby ensuring the correctness of the interpretations. Features present in an abductive explanation are those sufficient to imply the assigned risk category \cite{ignatiev2019abduction}, without redundancy. With them, we can improve the interpretability of the FRS model and understand how it makes its decisions. The features present in contrastive explanations are those that, if appropriately changed, would cause the patient's score to fall into a lower risk category (e.g., from moderate to low or from high to moderate) \cite{chou2022counterfactuals}.

This paper is organized as follows. In Section 1, we introduce the research problem and motivate the need for more transparent CVD risk assessments. Section 2 reviews the most relevant related work in explainable AI and risk‐prediction models. Section 3 describes our methodology in detail, outlining the logical framework, model formulation, and system architecture. Section 4 presents the experimental evaluation, implementation of the explainer, and an analysis of its performance. Finally, Section 5 summarizes our main findings, discusses their implications, and suggests directions for future work.

\section{Background}

\subsection{Cardiovascular diseases (CVDs)}
Cardiovascular disease (CVD) encompasses a broad range of conditions affecting the heart and blood vessels, and it remains one of the leading causes of death and disability worldwide. \cite{khan2021cardiovascular} This group of disorders includes ailments such as coronary artery disease, heart failure, and arrhythmias. The onset and progression of CVD are influenced by both non-modifiable factors, such as age and genetics, and modifiable factors, including lifestyle choices and environmental influences \cite{cimmino2023non}. 


\subsection{Framingham Risk Score}

The Framingham Risk Score is a tool that helps calculate the estimated risk of having cardiovascular disease in 10 years. This tool was created using the Framingham Heart Study, where the comprehensive data collected from the participants was analyzed to identify key risk factors. The study results were then translated into practical risk assessment tables by performing statistical analyses to determine how each risk factor contributes to the overall cardiovascular risk. These analyses produced regression coefficients for various features (such as age, cholesterol levels, blood pressure, and smoking status), which were then converted into point values. These point values are organized into tables that group the risk factors into different ranges, allowing clinicians to quickly assess an individual's risk level \cite{d2008general}.


The Framingham Risk Score is computed using dual-point tables, one for each feature for males and another for females. For a given individual, each feature is located within a specified range in the corresponding table, and the associated point value is added to the overall score. According to Table \ref{tab:cvd_points_male}, for instance, a would receive a total of 26 points.
\begin{table}[h]
    \scriptsize
    \centering
    \caption{CVD Prediction Points Table for Male adapted from
    \cite{d2008general}}
    \raggedright\footnotesize HDL: High-Density Lipoprotein, SBP: Systollic Blood Pressure
    \label{tab:cvd_points_male} 
  \renewcommand{\arraystretch}{1.2}
  \begin{tabularx}{\linewidth}{|c|*{7}{>{\centering\arraybackslash}X|}}
        \hline
        \textbf{Points} & \textbf{Age (years)} & \textbf{HDL (mg/dL)} & \textbf{Total Chol (mg/dL)} & \textbf{SBP - Not Treated (mm Hg)} & \textbf{SBP - Treated (mm Hg)} & \textbf{Smoker} & \textbf{Diabetic} \\
        \hline
        
        -2  &        & 60+     &         &    \textless 120     &         &        &        \\ \hline
        -1  &        & 50–59   &         &         &         &     &    \\ \hline
         0  & 30–34  & 45–49   & \textless160    & 120–129   & \textless120 & No & No       \\ \hline
         1  &        & 35–44 & 160–199 & 130–139   &   &        &        \\ \hline
         2  & 35–39  & \textless35 & 200–239 & 140–159 & 120–129 &        &        \\ \hline
         3  &        &         & 240–279 & 160+    & 130–139 &        & Yes    \\ \hline
         4  &        &         & 280+    &         & 140–159 &  Yes   &        \\ \hline
         5  & 40–44  &         &         &         & 160+    &     &        \\ \hline
         6  & 45+ &         &         &         &         &        &        \\ \hline
         8  & 50–54  &         &         &         &         &        &        \\ \hline
         10  & 55–59  &         &         &         &         &        &        \\ \hline
        11  & 60–64  &         &         &         &         &        &        \\ \hline
        12  & 65–69  &         &         &         &         &        &        \\ \hline
        14  & 70–74  &         &         &         &         &        &        \\ \hline
        15  & 75+    &         &         &         &         &        &        \\ \hline
    \end{tabularx}
    
\end{table}

Once the total score is calculated, it is mapped to Table \ref{tab:risk-points-male}, which determines the corresponding risk percentage. In our example, the man's risk would be >30\%. There are three possible risk categories: low-risk (0\% to 6\%), moderate-risk (6\% to 20\%), and high-risk (\(\geq 20\%\)) \cite{d2008general}. Therefore, the man would fall into the high-risk category.
\begin{table}[ht]
    \centering
    \caption{Risk percentage according to total points for male \cite{d2008general}}
    \label{tab:risk-points-male}
    \renewcommand{\arraystretch}{1.2}
    \begin{tabular}{|c|c|}
        \hline
        \textbf{Points} & \textbf{Risk (\%)} \\
        \hline
        $\leq -3$ or less & $<1$ \\
        \hline
        $-2$ & 1.1 \\
        \hline
        $-1$ & 1.4 \\
        \hline
        0 & 1.6 \\
        \hline
        1 & 1.9 \\
        \hline
        2 & 2.3 \\
        \hline
        3 & 2.8 \\
        \hline
        4 & 3.3 \\
        \hline
        5 & 3.9 \\
        \hline
        6 & 4.7 \\
        \hline
        7 & 5.6 \\
        \hline
        8 & 6.7 \\
        \hline
        9 & 7.9 \\
        \hline
        10 & 9.4 \\
        \hline
        11 & 11.2 \\
        \hline
        12 & 13.2 \\
        \hline
        13 & 15.6 \\
        \hline
        14 & 18.4 \\
        \hline
        15 & 21.6 \\
        \hline
        16 & 25.3 \\
        \hline
        17 & 29.4 \\
        \hline
        18+ & $>30$ \\
        \hline
    \end{tabular}
\end{table}

\subsection{First-order Logic over LRA}\label{subsec:logic}
In this work, we use first-order logic (FOL) to give explanations with guarantees of correctness. We use quantifier-free first-order formulas over the theory of linear rational arithmetic (LRA). Then, first-order variables are allowed to take values from the real numbers $\mathbb{R}$. For details, see \cite{kroening2016decision}. Therefore, we consider formulas as defined below:
\begin{equation}
        \begin{aligned}
             F, G &:= p \mid (F \wedge G) \mid (F \vee G) \mid (\neg F) \mid (F \to G),\\
             p &:= \sum^n_{i=1} w_i x_i \leq b, 
        \end{aligned}    
\end{equation}
such that $F$ and $G$ are quantifier-free first-order formulas over the theory of linear real arithmetic. Moreover, $p$ represents the atomic formulas such that $n \geq 1$, each $w_i$ and $b$ are fixed real numbers, and each $x_i$ is a first-order variable. For example, $(2.5x_1 + 3.1x_2 \geq 6) \wedge (x_1=1 \vee x_1=2) \wedge (x_1=2 \to x_2 \leq 1.1)$ is a formula by this definition. Observe that we allow standard abbreviations as $\neg (2.5x_1 + 3.1x_2 < 6)$ for $2.5x_1 + 3.1x_2 \geq 6$. 

Since we are assuming the semantics of formulas over the domain of real numbers, an \textit{assignment} $\mathcal{A}$ for a formula $F$ is a mapping from the first-order variables of $F$ to elements in the domain of real numbers. For instance, $\{x_1 \mapsto 2.3, x_2 \mapsto 1\}$ is an assignment for $(2.5x_1 + 3.1x_2 \geq 6) \wedge (x_1=1 \vee x_1=2) \wedge (x_1=2 \to x_2 \leq 1.1)$. An assignment $\mathcal{A}$ \textit{satisfies} a formula $F$ if $F$ is true under this assignment. For example, $\{x_1 \mapsto 2, x_2 \mapsto 1.05\}$ satisfies the formula in the above example, whereas $\{x_1 \mapsto 2.3, x_2 \mapsto 1\}$ does not satisfy it. Moreover, an assignment $\mathcal{A}$ \textit{satisfies} a set $\Gamma$ of formulas if all formulas in $\Gamma$ are true under $\mathcal{A}$.

A set of formulas $\Gamma$ is \textit{satisfiable} if there exists a satisfying assignment for $\Gamma$. To give an example, the set $\{(2.5x_1 + 3.1x_2 \geq 6), (x_1=1 \vee x_1=2), (x_1=2 \to x_2 \leq 1.1)\}$ is satisfiable since $\{x_1 \mapsto 2, x_2 \mapsto 1.05\}$ satisfies it. As another example, the set $\{(x_1 \geq 2), (x_1 < 1)\}$ is unsatisfiable since no assignment satisfies it. Given a set of formulas $\Gamma$ and a formula $G$, the notation $\Gamma \models G$ is used to denote \textit{logical consequence} or \textit{entailment}, i.e., each assignment that satisfies $\Gamma$ also satisfies $G$. As an illustrative example, let $\Gamma = \{x_1 = 2 , x_2 \geq 1\}$ and $G = (2.5x_1 + x_2 \geq 5) \wedge (x_1=1 \vee x_1=2)$. Then, $\Gamma \models G$. The essence of entailment lies in ensuring the correctness of the conclusion $G$ based on the set of premises $\Gamma$. In the context of computing explanations, as presented in \cite{ignatiev2019abduction}, logical consequence serves as a fundamental tool for guaranteeing correctness.
 
The relationship between satisfiability and entailment is a fundamental aspect of logic. It is widely known that, for all sets of formulas $\Gamma$ and all formulas $G$, it holds that $\Gamma \models G$ iff $\Gamma \cup \{\neg G\}$ is unsatisfiable. For instance, $\{x_1 = 2, x_2 \geq 1),  \neg((2.5x_1 + x_2 \geq 5) \wedge (x_1=1 \vee x_1=2))\}$ has no satisfying assignment since an assignment that satisfies $\{x_1 = 2 , x_2 \geq 1\}$ also satisfies $(2.5x_1 + x_2 \geq 5) \wedge (x_1=1 \vee x_1=2)$ and, therefore, does not satisfy $\neg((2.5x_1 + x_2 \geq 5) \wedge (x_1=1 \vee x_1=2))$. Since our approach builds upon the concept of logical consequence, we can leverage this connection in the context of computing explanations.

\subsection{Logic Based Explanations}
Due to the Framingham risk score being defined by assigning values to tables according to the features of a user, the scoring model can be readily adapted to a First-Order Logic (FOL) framework by formalizing the rules as logical implications. For example, for continuous variables such as age, one can express:
\[
40 \leq \text{age} < 45 \;\rightarrow\; \text{age\_points} = 5.
\]
For Boolean variables, such as smoking status (denoted by \(\text{is\_smoker}\)), the assignment of points is defined as follows: if \(\text{is\_smoker}\) is true, the patient receives \(4\) points; if false, \(0\) points are added. This can be represented in FOL as:
\[
(\text{is\_smoker} \rightarrow \text{smoker\_points} = 4)
\;\wedge\;
(\neg\text{is\_smoker} \rightarrow \text{smoker\_points} = 0).
\]
Once all features have been assigned their point values, the total score determines the patient’s risk category (low, moderate, or high). Explanations in this FOL framework then proceed by identifying two complementary types of feature-sets 

To compute the abductive features, we iterate over each feature in the current interpretation, remove it, and check whether the remaining features still logically entail the originally assigned risk category. If entailment still holds, that feature is deemed irrelevant and is not included in the abductive explanation \cite{ignatiev2019abduction}.

To compute the counterfactual features, we start from an empty set and incrementally add each mutable feature (i.e., excluding immutable features such as age or sex) together with the desired target category (e.g.\ “low risk”). If the conjunction of this set with the target category is unsatisfiable, we remove that feature from consideration. The resulting counterfactual explanation is the difference between the original interpretation and the final feature set \cite{chou2022counterfactuals}.

Applying this approach to the example presented previously, the 70-year-old man with diabetes, who is not on medication for systolic blood pressure, currently does not smoke and has a total cholesterol of 283 mg/dL, an HDL of 30 mg / dL and a systolic blood pressure of 170 mm Hg, we identify systolic blood pressure, diabetes, and age as abductive features. This occurs even though total cholesterol contributes more points than systolic blood pressure and diabetes, as shown in Table \ref{tab:cvd_points_male}. The fact that the Framingham Risk Score (FRS) is based on a point system may lead to the assumption that the features contributing the highest number of points are necessarily the most important. However, as this example illustrates, that is not always the case, highlighting the relevance of our proposed approach.

\section{Methodology}

\subsection{Data Collection}

To comprehensively evaluate our framework, we constructed a synthetic dataset that exhaustively covers all possible input combinations used in the FRS calculation. For each continuous feature, we selected representative values corresponding to the distinct ranges defined by the FRS guidelines. Since values within the same range yield equivalent risk contributions, this discretization permits a finite and complete enumeration of all meaningful input configurations.
\begin{table}[ht]
    \centering
        \caption{Number of possible values for each feature after continuous-features quantization for males and famale}
    \begin{tabular}{|c|c|c|}
        \hline
         \textbf{Feature} & \textbf{\#Values (Male)}  & \textbf{\#Values (Female)} \\ \hline
         Age & 10 & 10 \\ \hline
         HDL & 5 &  5\\ \hline
         Total ChoL & 5 & 5\\ \hline
         SBP & 5 & 6 \\ \hline
         Treatment for SBP & 2 &  2\\ \hline
         Smoker & 2 & 2 \\ \hline
         Diabetic & 2 &  2\\ \hline
    \end{tabular}
    \label{tab:male_features}
\end{table}

Based on Table~\ref{tab:male_features}, the total number of distinct FRS inputs is 22,000. This quantity is straightforward to process exhaustively. All distinct samples (each reflecting a legitimate combination of FRS input values) were automatically created using Python and the \texttt{pandas} package.  The relevant risk score for each sample was then calculated and explained by feeding this dataset into our logic-based explanation engine.


\subsection{Explainer}

Using the \texttt{z3py} API, we encoded the entire Framingham Risk Score (FRS) calculation for the created explanation as a set of logical constraints in the Z3 SMT solver, including threshold checks and risk table lookups.  Z3 uses logical inference to calculate the risk score based on the patient's input features.  Without using heuristics or approximation techniques, we were able to process our whole synthetic dataset of 22,000 samples and generate logically consistent explanations for each instance thanks to Z3's effective constraint-solving capabilities.  We produced two kinds of explanations using this formal encoding: counterfactual explanations, which identify the smallest changes to input features necessary to change the risk score to a given target value, and abductive explanations, which identify the minimal subset of input features sufficient to justify the computed risk score.

\subsection{Experiments}

The experiments in this study were conducted under two distinct scenarios. The first scenario aimed to identify abductive features, that is, those whose presence is sufficient to justify the assigned risk category. The goal was to enhance the interpretability of the FRS model and to gain a better understanding of the rationale behind its decisions. The second scenario focused on identifying counterfactual features, meaning variables that, if appropriately modified, could lead to a reclassification of the patient (e.g., from high to moderate or from moderate to low).

\section{Results}

In the first scenario, our explainer achieved goods results using the synthetic dataset with 22,000 inputs. The abductive explanations are richer, with nearly 77\% including five or more features, suggesting that justifying a given risk score typically requires citing a constellation of factors, where we first mensured the quantified how concise each explanation method tends to be. The Table \ref{tab:abductive_sparsity_by_method} shows this results.


\begin{table}[h]
  \centering
  \caption{Distribution of abductive explanation sparsity over all FRS inputs}
  \label{tab:abductive_sparsity_by_method}
  \begin{tabular}{|c|c|}
    \hline
    \textbf{Number of Features} & \textbf{Abductive (\%)} \\
    \hline
    3 & 4.00 \\ 
    4 & 18.14 \\ 
    5 & 25.15 \\ 
    6 & 35.97 \\ 
    7 & 16.05  \\ 
    8 & 0.70 \\ 
    \hline
  \end{tabular}
  \vspace{1ex}
\end{table}

Subsequently, we examined which features appear most often in each explanation type. To abductive explanations, age and systolic blood pressure dominate, appearing in over 90\% of justifications, as you can see in Table \ref{tab:presence_abductive}. Moreover, the distinction between modifiable factors (\textit{blood pressure, cholesterol, smoking status and medication})  and non‐modifiable factors (\textit{sex and age}) is clear: sex appears in only 30\% of cases, while all modifiable factors appear in 50–75\% of cases, highlighting their central role in justifying risk.

\begin{table}[ht]
\centering
\caption{Feature presence in abductive explanations over all inputs}
\label{tab:presence_abductive}
\begin{tabular}{|l|c|c|}
\hline
\textbf{Feature} & \textbf{Count} & \textbf{\% of all samples} \\
\hline
Age & 21\,593 & 98.2\% \\
Systolic blood pressure & 20\,329 & 92.4\% \\
Smoker status & 15\,662 & 71.2\% \\
HDL cholesterol & 14\,588 & 66.3\% \\
Total cholesterol & 13\,095 & 59.5\% \\
Treatment for SBP & 11\,257 & 51.2\% \\
Male sex & 6\,579 & 29.9\% \\
\hline
\end{tabular}
\end{table}

In the second scenario, displayed in Table \ref{tab:counterfactual_sparsity_by_method}, we observe that counterfactual explanations are sparse. Over 80\% involve at most two features (1‐feature: 47.17\%, 2‐features: 35.07\%), indicating that  only one or two features is enough to change the risk category.

\begin{table}[ht]
  \centering
  \caption{Distribution of counterfactual explanation sparsity over Moderate‐Risk and High‐Risk FRS inputs}
  \label{tab:counterfactual_sparsity_by_method}
  \begin{tabular}{|c|c|}
    \hline
    \textbf{Number of Features} & \textbf{Counterfactual (\%)} \\
    \hline
    1 & 47.17 \\
    2 & 35.07 \\
    3 & 13.06 \\
    4 & 3.32 \\
    5 & 0.54 \\
    6 & 0.84 \\
    \hline
  \end{tabular}
  \vspace{1ex}
\end{table}

In counterfactual explanations, \textit{systolic blood pressure} and \textit{total cholesterol} were the major factor to change the risk class, each one appearing in over 40\% of cases. The \textit{smoking status} and \textit{HDL cholesterol }are less frequent, suggesting that change blood pressure or cholesterol alone is often sufficient to cross risk thresholds. Table \ref{tab:presence_counterfactual} shows the results. Overall, these patterns align with clinical practice: abductive explanations highlight core drivers such as age and sex, while counterfactual explanations focus on modifiable factors for intervention.

\begin{table}[h]
\centering
\caption{Feature presence in counterfactual explanations (excluding low‐risk)}  
\label{tab:presence_counterfactual}
\begin{tabular}{|l|c|c|}
\hline
\textbf{Feature} & \textbf{Count} & \textbf{\%} \\ \hline
Systolic blood pressure & 8\,330  & 43.7\% \\
Total cholesterol & 8\,019 &  42.1\% \\
Treatment for SBP & 5\,958 & 31.3\% \\
HDL cholesterol & 4\,983 & 26.2\% \\
Smoker status & 2\,450 & 12.9\% \\
Sex, age & 0 & 0.0\% \\
\hline
\end{tabular}
\end{table}

Overall, these results align closely with established clinical practice: abductive explanations emphasize the foundational drivers of cardiovascular risk, while counterfactual explanations highlight the modifiable factors that clinicians target for intervention.

\section{Conclusion}

Two complementary explanation types (abductive and counterfactual) for the FRS were successfully produced by our system.  The abductive explanations demonstrate that our approach accurately reflects the established clinical hierarchy of risk factors.  Counterfactual explanations target specific areas for intervention to change the patient's risk category.
Future enhancements could include testing the explainer on real-world datasets and evaluating its effectiveness through expert assessment. Additionally, the approach can improve the counterfactual explanations to calculate specific features values to alter the risk classification.



\subsubsection{\ackname}
The authors acknowledge the support of Instituto Federal do Ceará (IFCE) through the research grant calls PIBITI No. 11/2024 and PIBIC No. 07/2024, issued by the PRPI/IFCE, as well as the support of Fundação Cearense de Apoio ao Desenvolvimento Científico e Tecnológico (FUNCAP) and Conselho Nacional de Desenvolvimento Científico e Tecnológico (CNPq) in the development of this work.




%
%
%
\bibliographystyle{splncs04nat}
\bibliography{references}

@inproceedings{ignatiev2019abduction,
  title={Abduction-based explanations for machine learning models},
  author={Ignatiev, Alexey and Narodytska, Nina and Marques-Silva, Joao},
  booktitle={33rd AAAI},
  year={2019}
}

@article{wilson1998prediction,
  title={Prediction of coronary heart disease using risk factor categories},
  author={Wilson, Peter WF and D’Agostino, Ralph B and Levy, Daniel and Belanger, Albert M and Silbershatz, Halit and Kannel, William B},
  journal={Circulation},
  volume={97},
  number={18},
  pages={1837--1847},
  year={1998},
  publisher={Lippincott Williams \& Wilkins}
}

@article{damen2016prediction,
  title={Prediction models for cardiovascular disease risk in the general population: systematic review},
  author={Damen, Johanna AAG and Hooft, Lotty and Schuit, Ewoud and Debray, Thomas PA and Collins, Gary S and Tzoulaki, Ioanna and Lassale, Camille M and Siontis, George CM and Chiocchia, Virginia and Roberts, Corran and others},
  journal={bmj},
  volume={353},
  year={2016},
  publisher={British Medical Journal Publishing Group}
}

@article{arnett20192019,
  title={2019 ACC/AHA guideline on the primary prevention of cardiovascular disease: a report of the American College of Cardiology/American Heart Association Task Force on Clinical Practice Guidelines},
  author={Arnett, Donna K and Blumenthal, Roger S and Albert, Michelle A and Buroker, Andrew B and Goldberger, Zachary D and Hahn, Ellen J and Himmelfarb, Cheryl Dennison and Khera, Amit and Lloyd-Jones, Donald and McEvoy, J William and others},
  journal={Journal of the American College of cardiology},
  volume={74},
  number={10},
  pages={e177--e232},
  year={2019},
  publisher={American College of Cardiology Foundation Washington, DC}
}

@article{kazi2024forecasting,
  title={Forecasting the economic burden of cardiovascular disease and stroke in the United States through 2050: a presidential advisory from the American Heart Association},
  author={Kazi, Dhruv S and Elkind, Mitchell SV and Deutsch, Anne and Dowd, William N and Heidenreich, Paul and Khavjou, Olga and Mark, Daniel and Mussolino, Michael E and Ovbiagele, Bruce and Patel, Sonali S and others},
  journal={Circulation},
  volume={150},
  number={4},
  pages={e89--e101},
  year={2024},
  publisher={Lippincott Williams \& Wilkins Hagerstown, MD}
}

@article{d2008general,
  title={General cardiovascular risk profile for use in primary care: the Framingham Heart Study},
  author={D’Agostino Sr, Ralph B and Vasan, Ramachandran S and Pencina, Michael J and Wolf, Philip A and Cobain, Mark and Massaro, Joseph M and Kannel, William B},
  journal={Circulation},
  volume={117},
  number={6},
  pages={743--753},
  year={2008},
  publisher={Lippincott Williams \& Wilkins}
}

@article{chou2022counterfactuals,
  title={Counterfactuals and causability in explainable artificial intelligence: Theory, algorithms, and applications},
  author={Chou, Yu-Liang and Moreira, Catarina and Bruza, Peter and Ouyang, Chun and Jorge, Joaquim},
  journal={Information Fusion},
  volume={81},
  pages={59--83},
  year={2022},
  publisher={Elsevier}
}

@article{chia2015validation,
  title={Validation of the Framingham general cardiovascular risk score in a multiethnic Asian population: a retrospective cohort study},
  author={Chia, Yook Chin and Gray, Sarah Yu Weng and Ching, Siew Mooi and Lim, Hooi Min and Chinna, Karuthan},
  journal={BMJ open},
  volume={5},
  number={5},
  pages={e007324},
  year={2015},
  publisher={British Medical Journal Publishing Group}
}

@article{wilson2005metabolic,
  title={Metabolic syndrome as a precursor of cardiovascular disease and type 2 diabetes mellitus},
  author={Wilson, Peter WF and D’Agostino, Ralph B and Parise, Helen and Sullivan, Lisa and Meigs, James B},
  journal={Circulation},
  volume={112},
  number={20},
  pages={3066--3072},
  year={2005},
  publisher={Lippincott Williams \& Wilkins}
}

@article{amiri2025cardiovascular,
  title={Cardiovascular disease risk prediction by Framingham risk score in women with polycystic ovary syndrome},
  author={Amiri, Mina and Mousavi, Maryam and Noroozzadeh, Mahsa and Azizi, Fereidoun and Ramezani Tehrani, Fahimeh},
  journal={Reproductive Biology and Endocrinology},
  volume={23},
  number={1},
  pages={1--11},
  year={2025},
  publisher={Springer}
}

@article{kasim2023validation,
  title={Validation of the general Framingham Risk Score (FRS), SCORE2, revised PCE and WHO CVD risk scores in an Asian population},
  author={Kasim, Sazzli Shahlan and Ibrahim, Nurulain and Malek, Sorayya and Ibrahim, Khairul Shafiq and Aziz, Muhammad Firdaus and Song, Cheen and Chia, Yook Chin and Ramli, Anis Safura and Negishi, Kazuaki and Nasir, Nafiza Mat},
  journal={The Lancet Regional Health--Western Pacific},
  volume={35},
  year={2023},
  publisher={Elsevier}
}

@article{precoma2019updated,
  title={Updated cardiovascular prevention guideline of the Brazilian Society of Cardiology-2019},
  author={Precoma, Dalton Bertolim and Oliveira, Gl{\'a}ucia Maria Moraes de and Simao, Antonio Felipe and Dutra, Oscar Pereira and Coelho, Otavio Rizzi and Izar, Maria Cristina de Oliveira and P{\'o}voa, Rui Manuel dos Santos and Giuliano, Isabela de Carlos Back and Alencar, Arist{\'o}teles Comte de and Machado, Carlos Alberto and others},
  journal={Arquivos brasileiros de cardiologia},
  volume={113},
  pages={787--891},
  year={2019},
  publisher={SciELO Brasil}
}

@article{khan2021cardiovascular,
  title={Cardiovascular diseases},
  author={Khan, T},
  journal={World Health Organization},
  year={2021}
}

@article{cimmino2023non,
  title={Non-conventional risk factors:“fact” or “fake” in cardiovascular disease prevention?},
  author={Cimmino, Giovanni and Natale, Francesco and Alfieri, Roberta and Cante, Luigi and Covino, Simona and Franzese, Rosa and Limatola, Mirella and Marotta, Luigi and Molinari, Riccardo and Mollo, Noemi and others},
  journal={Biomedicines},
  volume={11},
  number={9},
  pages={2353},
  year={2023},
  publisher={MDPI}
}

@article{kroening2016decision,
  title={Decision procedures for propositional logic},
  author={Kroening, Daniel and Strichman, Ofer and Kroening, Daniel and Strichman, Ofer},
  journal={Decision Procedures: An Algorithmic Point of View},
  pages={27--58},
  year={2016},
  publisher={Springer}
}

\end{document}